\documentclass[twocolumn]{article}
\usepackage[top=2cm, bottom=2.5cm, left=2cm, right=2cm]{geometry}
\usepackage{times,textcomp}
\usepackage[T1]{fontenc}
\usepackage[utf8]{inputenc}
\usepackage{amssymb,amsmath}
\usepackage{enumitem}
\usepackage{eurosym}
\usepackage{subcaption}

\usepackage{titlesec}

\usepackage{titlesec} 
\titleformat{\section}{\large\bfseries}{\thesection}{1em}{\MakeUppercase} 
\titleformat{\subsection}{\large\bfseries}{\thesubsection}{1em}{}
\titlespacing*{\section}{0pt}{\parskip}{0pt}
\titlespacing*{\subsection}{0pt}{\parskip}{0pt}

\usepackage{longtable,booktabs}
\usepackage{graphicx,grffile}
\usepackage{fancyhdr}

\title{\huge\bfseries Financial Time Series Data Processing for Machine Learning}

\usepackage{authblk}

\author{Fabrice Daniel}

\affil{\small Artificial Intelligence Department of Lusis, Paris, France\\fabrice.daniel@lusis.fr\\http://www.lusis.fr}

\date{June 2019}
\begin{document}

\maketitle

\begin{abstract}
This article studies the financial time series data processing for machine learning. 
It introduces the most frequent scaling methods, then compares the resulting stationarity 
and preservation of useful information for trend forecasting. It proposes an empirical test
based on the capability to learn simple data relationship with simple models.
It also speaks about the data split method specific to time series, 
avoiding unwanted overfitting and proposes various labelling for classification and regression.
\end{abstract}


\section{Introduction}

In the field of machine learning Time Series are very special data needed their specific processing and methods\cite{IntroTSF}\cite{Forecasting}.

On top of that, Financial data adds a big challenge due to their proportion of randomness and their 
non-stationary nature\cite{analysisTS}\cite{statFinTS}.

There is a lot of research relative to the Financial Market forecast with Machine Learning\cite{AFML}.

However, many studies only cover one type of data scaling or labelling while the 
decisions made on this step can have a huge impact on the results. 
Not only in term of pure model performances metrics but in term of capabilities to really implement
a profitable trading strategy based on the model.

This study covers the following points:

\begin{itemize}
	\item Pre-processing and Stationarity
	\item Pre-processing and preservation of useful prices relationships
	\item Labelling for classifiers and regressors
\end{itemize}

\section{Stationarity}

Before to work on any price forecast model we need to pre-process our 
historical prices then we have to make sure the resulting data are stationary. 

We evaluate three of the most frequent pre-processing, starting by the
price returns, then two scaling methods: \textit{MinMax} and \textit{Standardization}.

For this purpose, we use the SPY daily closing prices between 1993 and 2019.

\begin{figure}[!h]
	\caption{SPY daily closing prices}
	\centering
	  \includegraphics[width=0.45\textwidth]{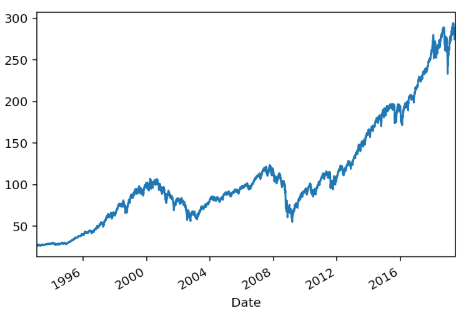}
	  \label{fig:spy}
\end{figure}

Let's first apply an Augmented Dickey-Fuller test\cite{DF}\cite{introStatTS} to the raw data as reference.

For such a sample size, the ADF of a dataset with trend must
be below -3.96 to reject the null hypothesis with a 1\% confidence \cite{ADF}.

\begin{itemize}
    \item ADF Statistic: -0.226901
    \item p-value: 0.991042
\end{itemize}

As expected with an ADF of -0.22 p-value of 0.99 the process is not stationary.

The most common way to make a time series stationary is differencing.
In the case of financial data we can simply compute the returns.

\begin{figure}[!h]
	\caption{SPY daily returns}
	\centering
	  \includegraphics[width=0.45\textwidth]{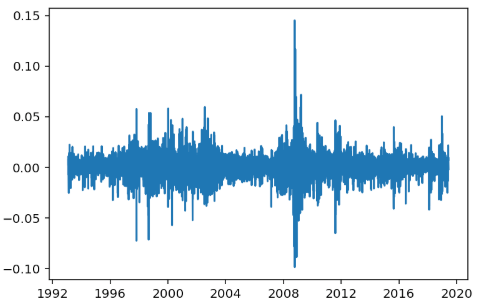}
\end{figure}

So, when applied on returns the ADF test gives:

\begin{itemize}
    \item ADF Statistic: -14.954070
    \item p-value: 0.000000
\end{itemize}

This confirms the time series of returns is stationary.

\section{Scaling}\label{chapter:scaling}

The returns data could be a good baseline, especially because the resulting very small 
values, near from 0 makes them directly compatible with a deep learning approach. 

While this is very suitable for a single feature time series, 
if we use multiple features like the High/Low prices, the Volume or any technical indicator
we can face an issue by simply using such a return-based method as it does not preserve the information
of the position of a feature relative to each other.

In this situations some other scaling methods must be used like \textit{MinMax} or \textit{Standardization}.
They are widely used in Machine Learning and enable to keep the relative position of each feature.

Before to evaluate these scaling methods, let's first introduce another aspect of 
data processing for time series in Machine Learning context, the slicing.

When doing Machine Learning on Financial Time Series, the model generally takes a time window 
as input, for instance 20 consecutive closing prices. The number of prices
used for this time window is defined as the \textit{lookback period}.

A frequent label found in many papers is the next price change, but we will see later this can be more sophisticated.

So let's assume a time series of $T$ consecutive stock returns $\{r_0,\ldots,r_{T-1}\}$

Building a training set $S$ consists on creating as series of $K$ slices $S = \{S_0,\ldots,S_{K-1}\}$, each of size $n > 1$, 
where $S_t = \{r_{t-n},\ldots, r_{t-1}\}$. 

For a model predicting the next return the label is defined by $y_{t-n}=r_t$.

Each $S_t$ slice is created by incrementing $t$ by steps of 1 or more. As an example, for 
an increment of 1, and a slice size of 20 returns, the first two sets are: 

$$(S_0=\{r_0,\ldots,r_{19}\},y_0=r_{20})$$
$$(S_1=\{r_1,\ldots,r_{20}\},y_1=r_{21})$$

After the slicing was done, we have $K$ slices than can be scaled independently from each other. 

We expect our Machine Learning model to identify price patterns 
leading to up or down move. Scaling each slice independently can make the training easier 
by removing the global range effect due to the long-term market trend.

\begin{figure}[!h]
  \centering
  \caption{Compare scale then slice, with slice then scale}\label{fig:slices-scaling}
  \begin{subfigure}[b]{0.2\textwidth}
	  \includegraphics[width=\textwidth]{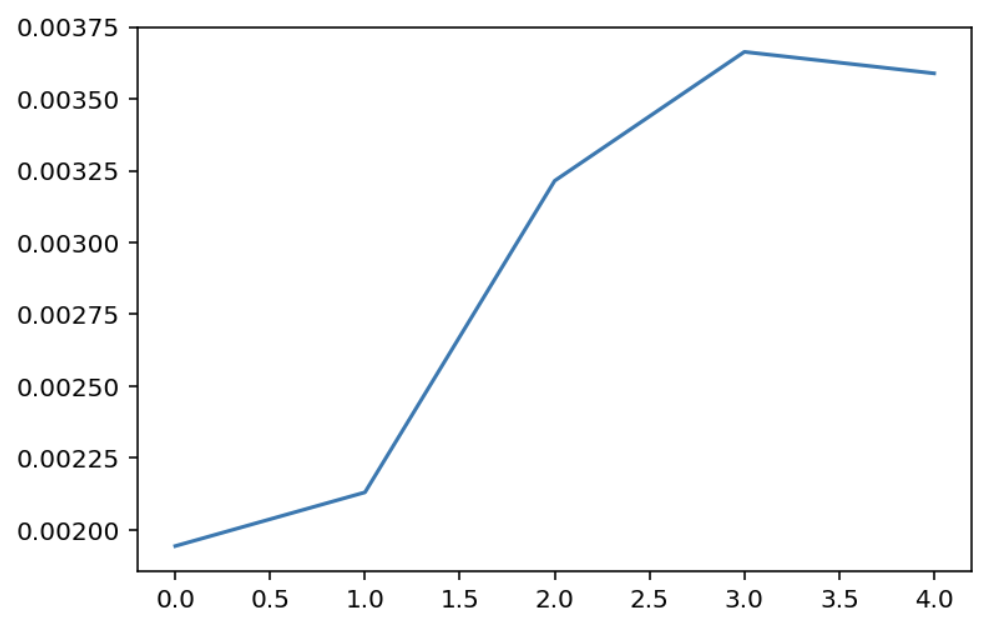}
	  \caption{First slice}
	  \label{fig:scale-slice-0}
  \end{subfigure}
  ~ 
  \begin{subfigure}[b]{0.2\textwidth}
	  \includegraphics[width=\textwidth]{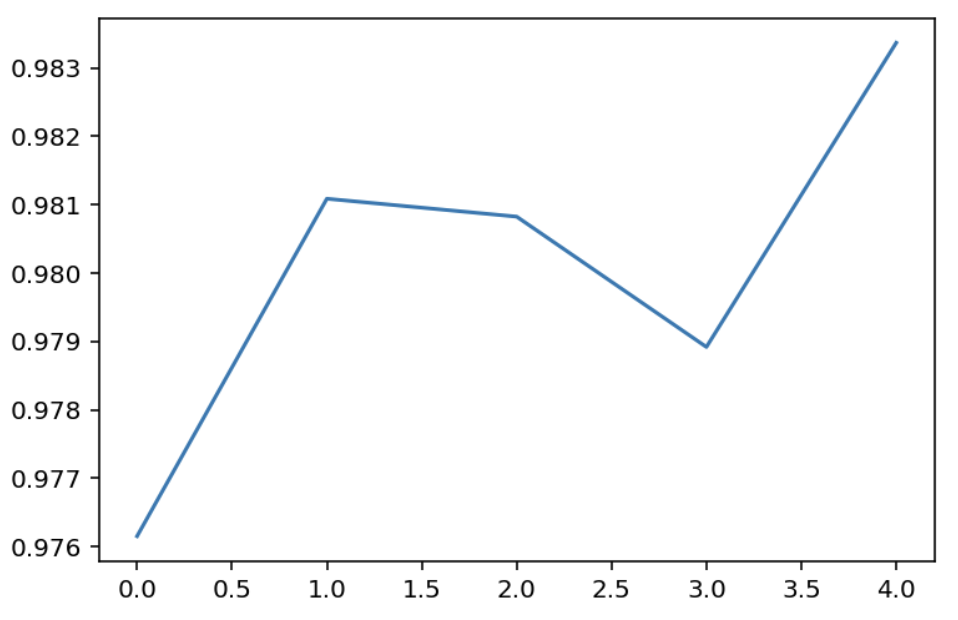}
	  \caption{Last slice}
	  \label{fig:scale-slice-1}
  \end{subfigure}
  \begin{subfigure}[b]{0.2\textwidth}
	\includegraphics[width=\textwidth]{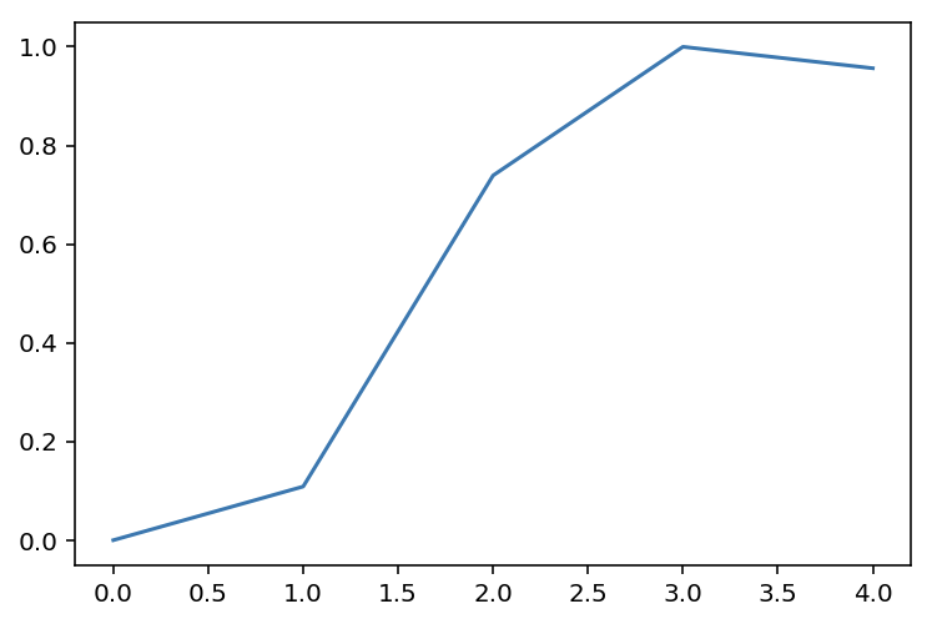}
	\caption{First slice}
	\label{fig:slice-scale-0}
  \end{subfigure}
  ~ 
  \begin{subfigure}[b]{0.2\textwidth}
	\includegraphics[width=\textwidth]{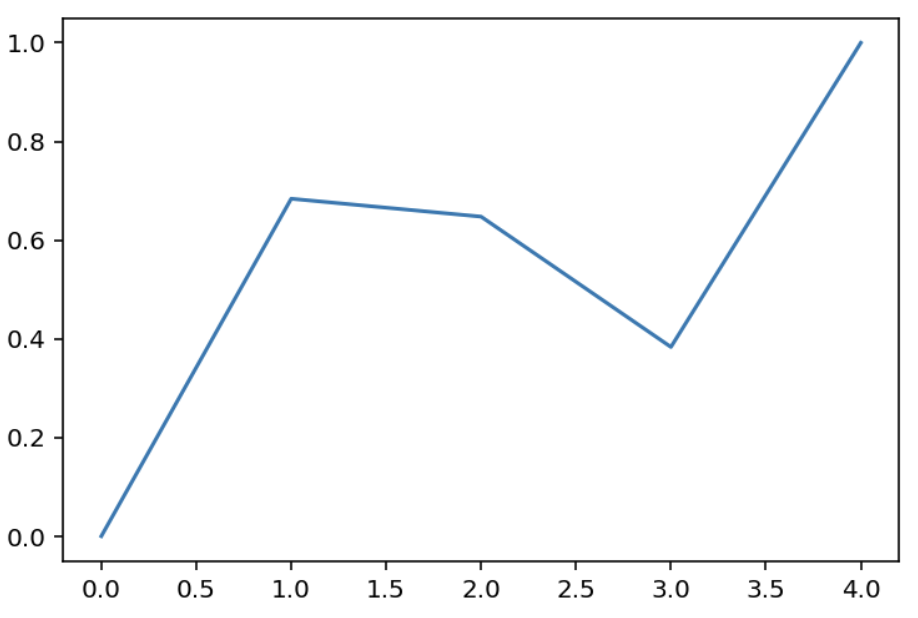}
	\caption{Last slice}
	\label{fig:slice-scale-1}
  \end{subfigure}
\end{figure}

For instance, here are the first and last slices when we scale first with a \textit{minmax} then slice 
(Figures \ref{fig:scale-slice-0}, \ref{fig:scale-slice-1})
and when we slice first then scale (Figures \ref{fig:slice-scale-0}, \ref{fig:slice-scale-1}).

It clearly appears Figure \ref{fig:slices-scaling} that slicing first then scaling 
enables getting each training example being within the same range, so making 
easier the model training.

Now let's find the best scaling method in a more formal way. 

For this purpose, we scale the training set then do the following:

\begin{itemize}
	\item Check stationarity with ADF
	\item Check if the information is preserved for machine learning context by training a 
	simple model identifying simple prices relationship into slices
\end{itemize}

We assume that if the model cannot learn a very simple price relationship inside each slice, like for 
instance identifying if the last close of the slice is above the close 5 bars ago, then 
there is nearly no chance for the model to be able to learn any price pattern leading to 
the future price changes.

If a simple price relationship is preserved, we expect the model to learn it nearly perfectly. 

So, the scaling method we select is the one enabling a model to learn
simple prices relationship inside a slice with the best efficiency.

We test two common scaling methods, the \textit{MinMax} and the \textit{Standardization}.

\textbf{MinMax}

Scale each slice into a $[0,1]$ or $[-1,1]$ range

Assume $x_{min}$ and $x_{max}$ the smallest and highest $x$ values.
And assume $min$ and $max$ the feature range, so $[0,1]$ for instance.

$$z = \frac{x-x_{min}}{x_{max}-x_{min}}(max - min) + min$$

\textbf{Standardization}

Scale each slice by removing the mean and scaling to unit variance.

$$z = \frac{x-\mu}{\sigma}$$

\begin{itemize}
	\item $\mu$ : mean
	\item $\sigma$ : standard deviation
\end{itemize}

After scaling and reshaping the SPY prices we apply an ADF test.

\begin{table}[!h]
\centering	
\begin{tabular}{|l|r|r|}
	\hline
	Scaling & ADF & p-value \\
	\hline	
	MinMax & -41.72 & 0.000 \\
	Standardization & -56.90 & 0.000 \\	
	\hline
\end{tabular}
\caption{ADF Test per scaling}\label{table:ads-test}
\end{table}

Each Scaling method results in a non-stationary dataset.

If prices relationship is preserved, we assume a simple model can be trained to identify
conditions like: 

\begin{itemize}
	\item $C_t > C_{t-5}$\footnote{$C_t$: closing price at time $t$, $C_{t-5}$ closing price 5 periods (days) ago}
	\item $C_t > EMA5_t$\footnote{EMA5 : 5 periods Exponential Moving Average}
	\item $C_t > HC10_t$\footnote{HC10: Highest close of the 10 last bars}
\end{itemize}

The third one is the most complex as the highest close over the past 10 bars 
is not always at the same position.

We create a binary label for each of these conditions then train 
the following LSTM model.

\begin{table}[!h]
	\centering	
	\begin{tabular}{|l|r|}
		\hline
	    Lookback (Slices) & 20 bars \\
		Features & Close  \\
		1 LSTM &  64 units (tanh) \\
		1 Dense output &  2 units (softmax) \\
		Bias & No \\
		Dropout & No \\
		Recurrent Dropout & No \\
		Optimizer & Adam \\
		Epochs & 100 \\
		Batch Size & 64 \\
		Training/Validation & 80/20 \\
		\hline
	\end{tabular}
	\caption{Model Detail and Hyperparameters}\label{table:ads-test}
\end{table}

Using 2 units output with softmax instead of a single unit binary output with sigmoid 
enables making the models more generic when increasing the number of classes. 

\begin{figure}[!h]
	\centering
	\caption{Loss of each model per scaling}\label{fig:slices-scaling}
	\begin{subfigure}[b]{0.2\textwidth}
		\includegraphics[width=\textwidth]{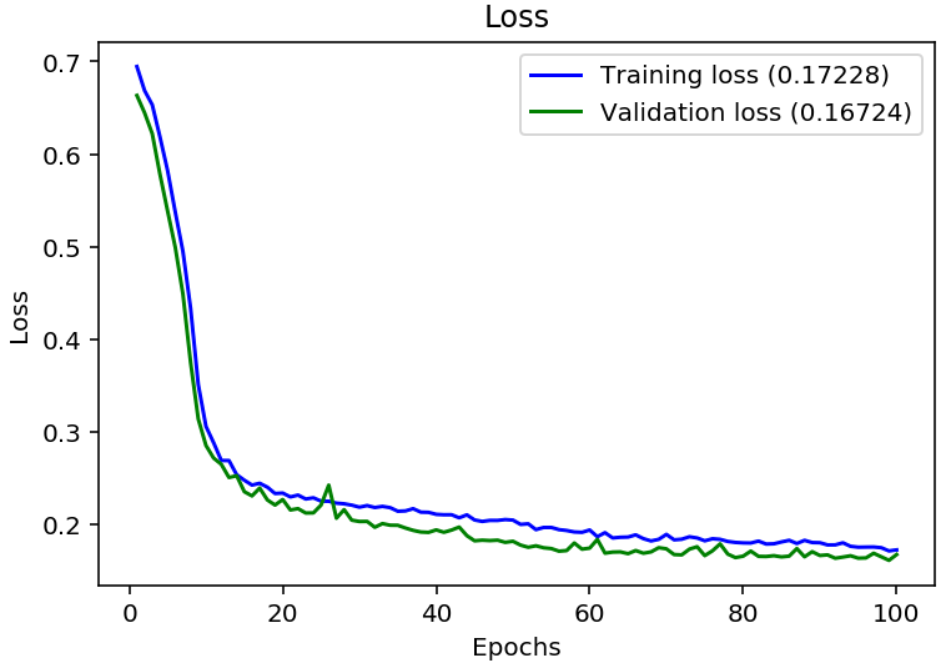}
		\caption{$C_t > C_{t-5}$ minmax}
		\label{fig:c-c5-minmax}
	\end{subfigure}
	~ 
	\begin{subfigure}[b]{0.2\textwidth}
		\includegraphics[width=\textwidth]{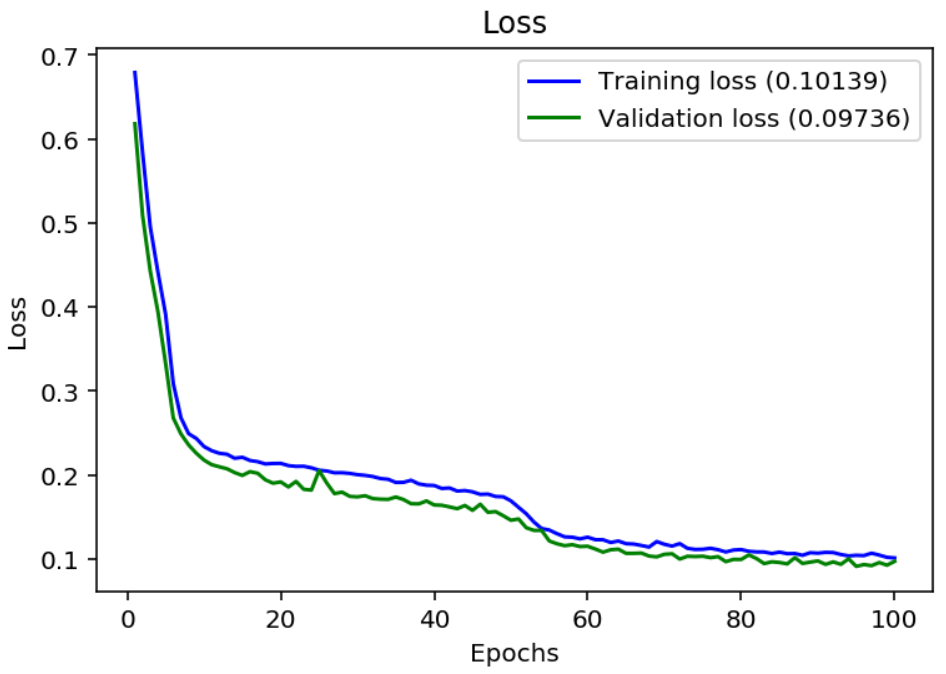}
		\caption{$C_t > C_{t-5}$ std}
		\label{fig:c-c5-std}
	\end{subfigure}

	\begin{subfigure}[b]{0.2\textwidth}
	  \includegraphics[width=\textwidth]{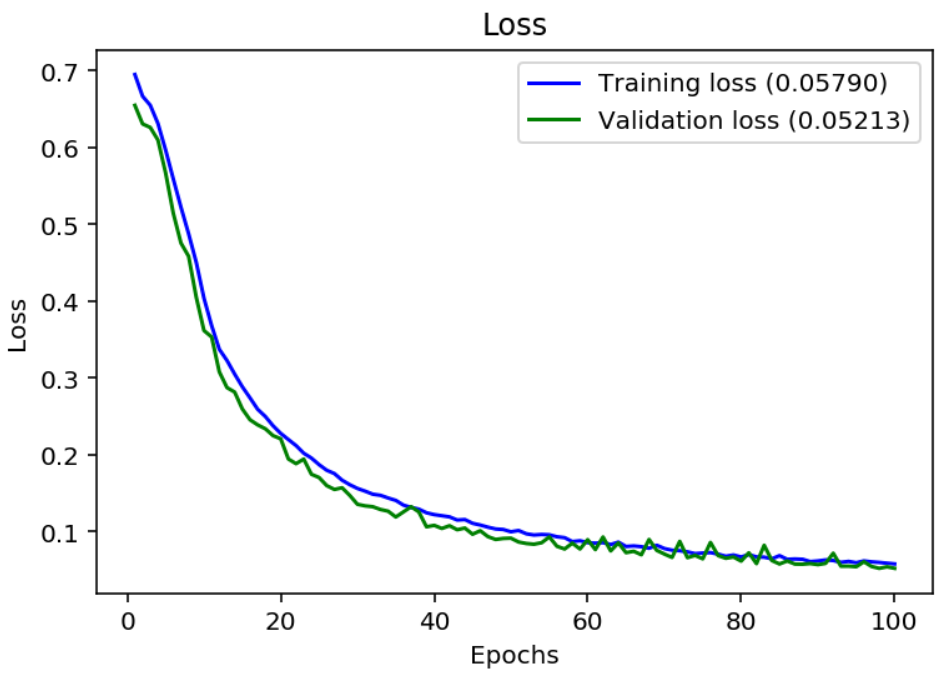}
	  \caption{$C_t > EMA5_t$ minmax}
	  \label{fig:c-ema5-minmax}
	\end{subfigure}
	~ 
	\begin{subfigure}[b]{0.2\textwidth}
	  \includegraphics[width=\textwidth]{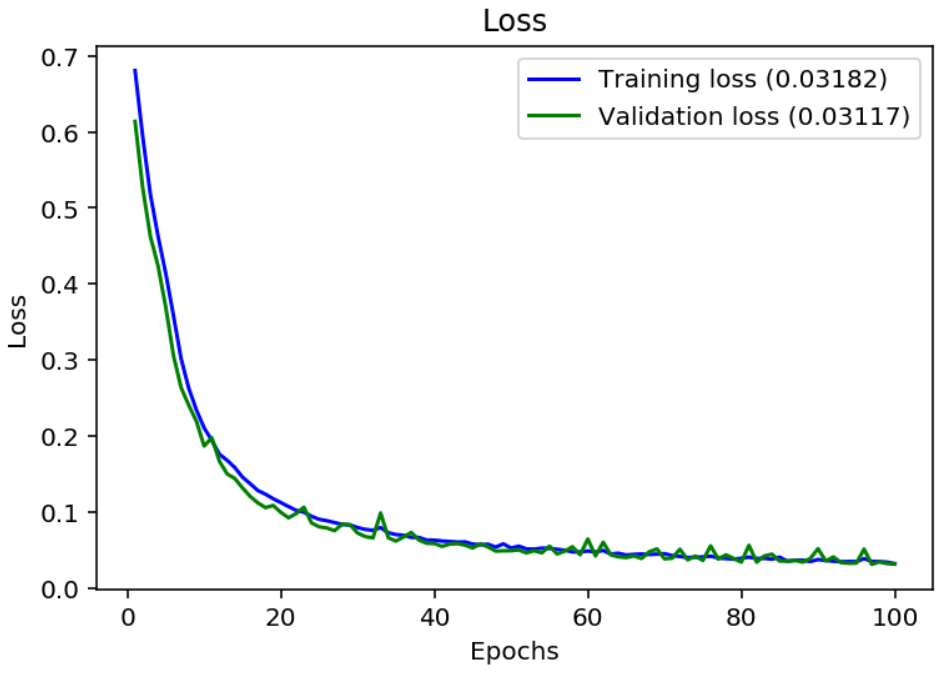}
	  \caption{$C_t > EMA5_t$ std}
	  \label{fig:c-ema5-std}
	\end{subfigure}

	\begin{subfigure}[b]{0.2\textwidth}
		\includegraphics[width=\textwidth]{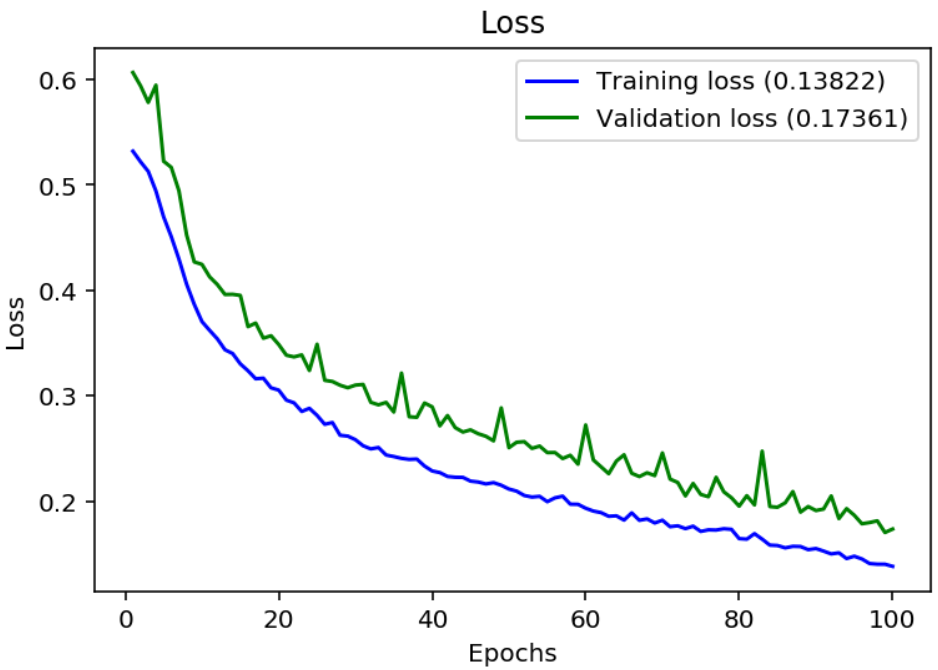}
		\caption{$C_t > HC10_t$ minmax}
		\label{fig:c-hc10-minmax}
	  \end{subfigure}
	  ~ 
	  \begin{subfigure}[b]{0.2\textwidth}
		\includegraphics[width=\textwidth]{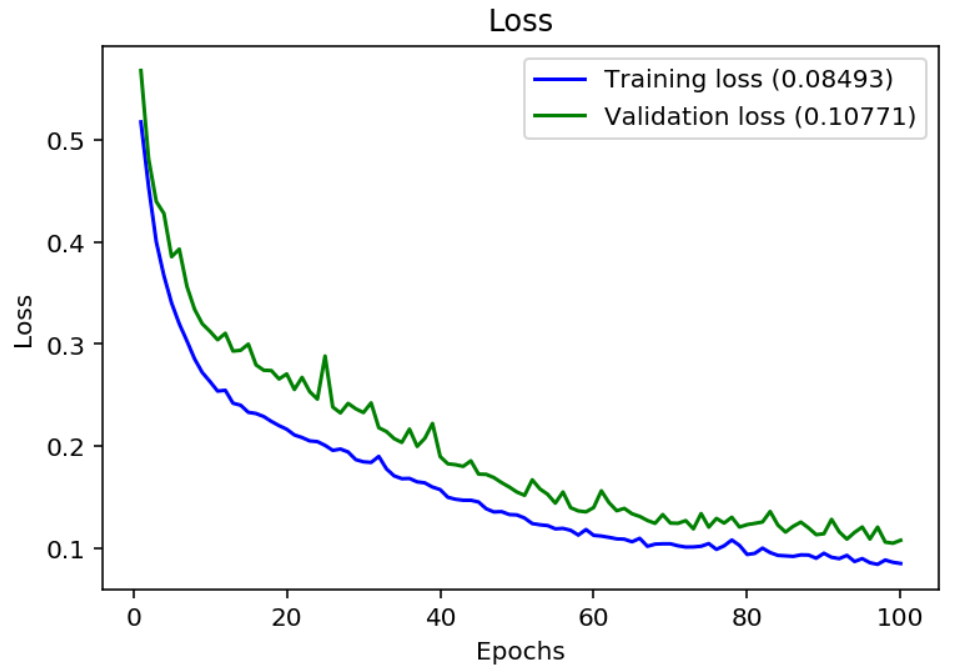}
		\caption{$C_t > HC10_t$ std}
		\label{fig:c-hc10-std}
	  \end{subfigure}

  \end{figure}

\newpage

The precisions on validation set are the following:

\begin{table}[!h]
\centering
\begin{tabular}{|l|r|r|}
	\hline
	Scaling & MinMax & Standardization \\
	\hline	
	$C_t > C_{t-5}$ & 0.920 & 0.978 \\
	$C_t > EMA5_t$ & 0.993 & 0.998 \\	
	$C_t > HC10_t$ & 0.910 & 0.904 \\	
	\hline
\end{tabular}
\caption{Precision of each model per scaling}\label{table:prec-scaling}
\end{table}

The two first cases are learned with a nearly perfect generalization. 

The last case shows that validation loss is slightly not as good as the training 
but seems to be able to continue to improve if we increase the number of epochs. 

By training the same LSTM on 150 epochs with 0.2 dropout and 
0.2 recurrent dropout to improve the generalization, we get the following loss.

\begin{figure}[!h]
	\centering
	\caption{HC10 model on 150 Epochs with dropouts}\label{fig:slices-scaling}
	\begin{subfigure}[b]{0.2\textwidth}
		\includegraphics[width=\textwidth]{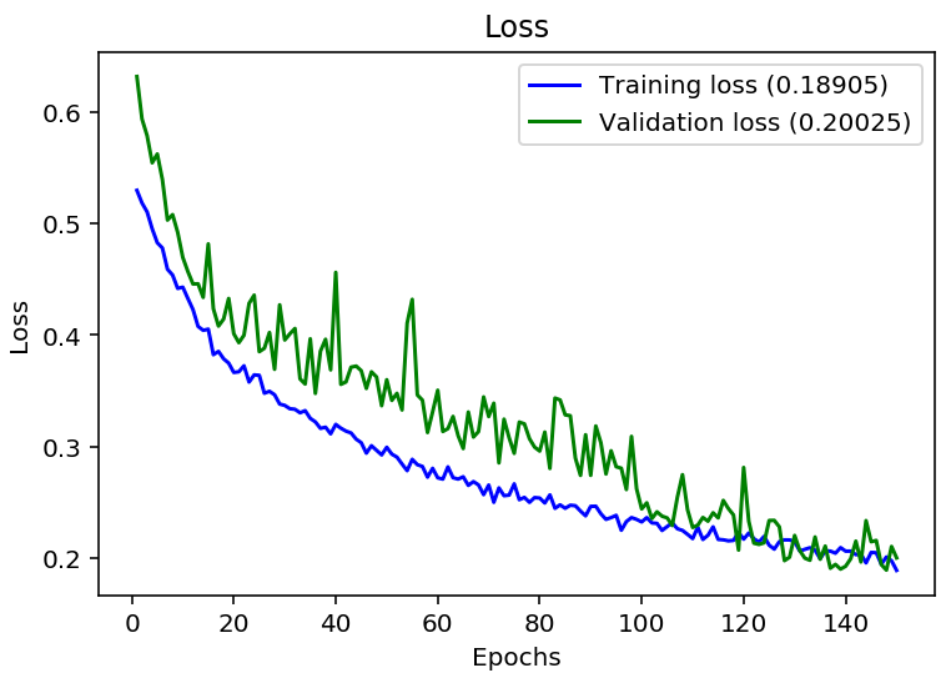}
		\caption{$C_t > HC10_t$ minmax}
		\label{fig:c-c5-minmax}
	\end{subfigure}
	~ 
	\begin{subfigure}[b]{0.2\textwidth}
		\includegraphics[width=\textwidth]{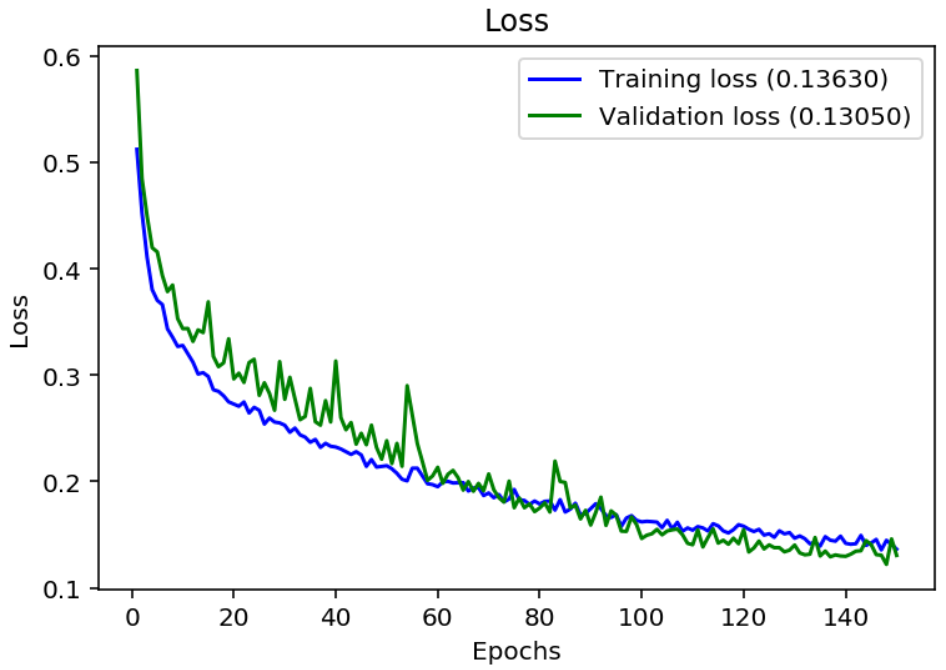}
		\caption{$C_t > HC10_t$ std}
		\label{fig:c-c5-std}
	\end{subfigure}
\end{figure}

The precisions on validation set are the following:
\begin{table}[!h]
\centering
\begin{tabular}{|l|r|r|}
	\hline
	Scaling & MinMax & Standardization \\
	\hline		
	$C_t > HC10_t$ & 0.970 & 0.968 \\	
	\hline
\end{tabular}
\caption{HC10 Precision per scaling}\label{table:prec-150}
\end{table}

These scaling methods preserve the prices relationship. It's also the case if we add new features like High, Low, or any 
overlaid indicator\footnote{An indicator plotted to the same chart than the prices (e.g. moving averages, Bollinger Bands)}. 
So, for instance the relative position of a closing price and a moving average is preserved.

The two scaling methods gives similar results but \textit{Standardization} looks a little bit better , 
especially when looking at the validation loss shape in the last case.

\section{Data Split}

In machine learning, splitting data into a training, validation and test set 
is often performed by using the following process:

\begin{enumerate}
	\item Shuffle the full data set
	\item Split into training/validation/test sets
\end{enumerate}

On time series this process leads to a risk of partially fitting to validation and test set during training
so getting very good metrics while getting poor results when applying the resulting trading strategy to real-time data. 

Remind the training set $S = \{S_0,\ldots,S_{K-1}\}$ has $K$ overlapped slices $S_t = \{r_{t-n},\ldots, r_{t-1}\}$.

Assume we have these two first slices:

$$(S_0=\{r_0,\ldots,r_{19}\},y_0=r_{20})$$
$$(S_1=\{r_1,\ldots,r_{20}\},y_1=r_{21})$$

If we shuffle before to split, we can have $S_0$ in the training set while $S_1$ can be part of the validation set.

In this case $S_0$ and $S_1$ have $\{r_1,\ldots,r_{19}\}$, so $95\%$ of their data, in common. 

If the shuffle is uniform, the biggest proportion of data in the validation set slices will also be part of the 
training set. In the example above for instance, up to 95\% of the validation set data can be part of the training set slices.

So, depending on the model type, while fitting the training set, we also have a chance to partially fit the validation set. 
This can result in unreliable validation loss showing good results without overfitting while its real metrics 
on very new data taken independently can be poor with a strong overfitting.

The proper way to split and shuffle the training set for Financial Time Series is the following:

\begin{enumerate}
	\item Split into training/validation/test sets \\
	
	The split is performed over $S$ such as:
	\begin{align}
		train &=\{S_0,\ldots,S_{ts-1}\} \\
		val   &=\{S_{ts},\ldots,S_{ts+vs-1}\} \\
		test  &=\{S_{ts+vs},\ldots,S_{K-1}\} 				
	\end{align}

	Where:

	\begin{itemize}
		\item $ts$ : training set size
		\item $vs$ : validation set size
		\item $K$ : total number of slices
	\end{itemize}

	\item Shuffle the training set
	
	So only $train = \{S_0,\ldots,S_{ts-1}\}$ is shuffled.

\end{enumerate}
 
\newpage

\section{Features}

In Financial time series the raw features are the Open, High, Low, Close and the Volume\footnote{As an OTC Market, the Forex does not includes the Volume. When present, it's only a synthetic indicator
built from the number of ticks during the period. Many people use it as a proxy for the market activity.}.

To take a decision, traders often use technical indicators calculated from these features. 
These indicators can be used as additional features to help the model.

When used as input in a Neural Network model, these indicators must be scaled 
but the method to use differs depending on their nature. The overlaid indicators must be scaled
but the separated indicators are generally moving in a narrow range of values, no matter if the 
price of the instrument is around 10 or 1000, so they need to be scaled separately with specific methods.

The overlaid indicators must be scaled all together with the prices to preserve their relationship. For instance
this enable the model to use the relation between the closing price and one or several indicators 
like moving averages.

The Volume and the separated indicators must have their own independent scaling. 

The not bounded indicators must be scaled with \textit{MinMax} or \textit{Standardization}. 

The bounded indicators can benefit from being divided by their maximum value (100 for the RSI). 
This preserve fixed values that can make sense for traders, like the overbought/oversold levels.
For instance, 70/30 for the RSI becomes 0.7/0.3 in the scaled version.

Note that when using another type of model like Random Forest or XGBoost, the scaling 
is not always required. Everything depends on the nature of the data. For instance
the RSI is bounded between 0 and 100, so it can directly be used in a Random Forest without any scaling.

When using a single feature, like the closing prices, the input shape is simply defined by $(m,s)$

When using multiple features, like the high, low, close and volume, the input shape becomes $(m,s,i)$

\begin{itemize}
    \item $m$ : samples
    \item $s$ : timesteps
    \item $i$ : features
\end{itemize}

If we build a training data set with 1000 slices of 20 bars, 
each with open, high, low, close and volume, the training set shape is : $(1000,20,5)$

When a model does not accept three dimensional inputs, a reshape could be necessary. 
Making the previous example flattened with a shape of $(1000,100)$

\section{Labeling}

Labelling can have a major influence on a model training to Financial Time Series. Some labels
can also results in non-tradable strategies. For instance, predicting a technical indicator created from lagged 
prices can be not usable for real trading, even when getting very accurate predictions.

Models can be of two type, classifier or regressor. A classifier generally attempts to predict 
the probability for the market to go Up or Down for a given time horizon while a regressor tries to predict the future price.

By far, the most frequent label found in the literature is the next return. However, many other labels can be 
used.

Table \ref{table:labeling} shows some examples of the labels we often use in our research at Lusis.

\begin{table}[!h]
	\centering
	\begin{tabular}{|l|l|}
		\hline
		Label & Description \\
		\hline	
		N bars Up/Down  & Classifier on $C_{t+n} > C_t$ \\
		N bars price change & Regressor on $C_{t+n} - C_t$  \\	
		N bars log returns & Regressor on $log(\frac{C_{t+n}}{C_t})$  \\	
		N bars Moving Average & Classifier on $MA_{t+n} > MA_t$  \\	
		N bars trend Strength & Regressor on Trend  \\
		N bars trend Direction & Classifier on Trend  \\
		\%Q after N bars & Regressor on \%Q \\
		QClass after N bars & Classifier on QClass \\
		\hline
	\end{tabular}
	\caption{Labels often used in our research}\label{table:labeling}
\end{table}

\%Q\footnote{\textit{Q} stands for \textit{Quality}} is a specific metric we especially created for focusing on the most tradable models. 

If $C_t$ is the closing price at time $t$, assume the corresponding time series slice of size $m$ ending by $C_t$ is defined by $S_t={C_{t-m+1},\ldots,C_t}$

Then

$$\%Q_{t+1}^{t+n}=\frac{HH_{t+1}^{t+n} - C_t}{HH_{t+1}^{t+n} - LL_{t+1}^{t+n}}$$

Where

\begin{itemize}
	\item $n$ : time horizon in number of bars
	\item $\%Q_{t+1}^{t+n}$ : \%Q between $t+1$ and $t+n$
	\item $HH_{t+1}^{t+n}$ : Highest High price between $t+1$ and $t+n$
	\item $LL_{t+1}^{t+n}$ : Lowest Low price between $t+1$ and $t+n$
\end{itemize}

So \%Q is interpreted as following:

\begin{itemize}
	\item $\%Q = 1$ when we have a perfect up move without any drawdown during the next $n$ bars
	\item $\%Q = 0$ when we have a perfect down move without any drawup during the next $n$ bars
	\item $\%Q = 0.5$ when we have an equally Up and Down move during the next $n$ bars
\end{itemize}

So the nearest from 1 or 0 \%Q is, the more tradable the prediction is. It corresponds to 
conditions where the MAE\footnote{Maximum Adverse Excursion} 
is lower than the MFE\footnote{Maximum Favorable Excursion} between $t$ and $t+n$.

\%Q corresponds to a risk/reward ratio. For instance, $\%Q = .75$ is a 1:3 risk/reward. 

The QClass label derivates from \%Q by using thresholds as class separators. 

\begin{table}[!h]
	\centering
	\begin{tabular}{|l|l|l|}
		\hline
		Class & Condition & Meaning\\
		\hline	
		0  & $\%Q >= 0.6$ & Up \\
		1 &  $0.4 < \%Q < 0.6$ & Neutral \\	
		2 & $\%Q <= 0.4$ & Down  \\	
		\hline
	\end{tabular}
	\caption{QClass example}\label{table:qclass}
\end{table}

The \textit{Trend Strength} and \textit{Trend Direction} labeling can be built by different ways. 
It can, for instance, be based on the shape of a linear regression, 
or on the percentage of closing prices above a simple moving average\footnote{When prices are rising up, a majority of them
are above their moving average}.

\section{Long term trend bias}

Presence of a Long Term Up Trend on stock market can make the model training 
to fit the Upside moves only. 

\begin{figure}[!h]
	\caption{Confusion Matrix on Validation Set}
	\centering
	  \includegraphics[width=0.45\textwidth]{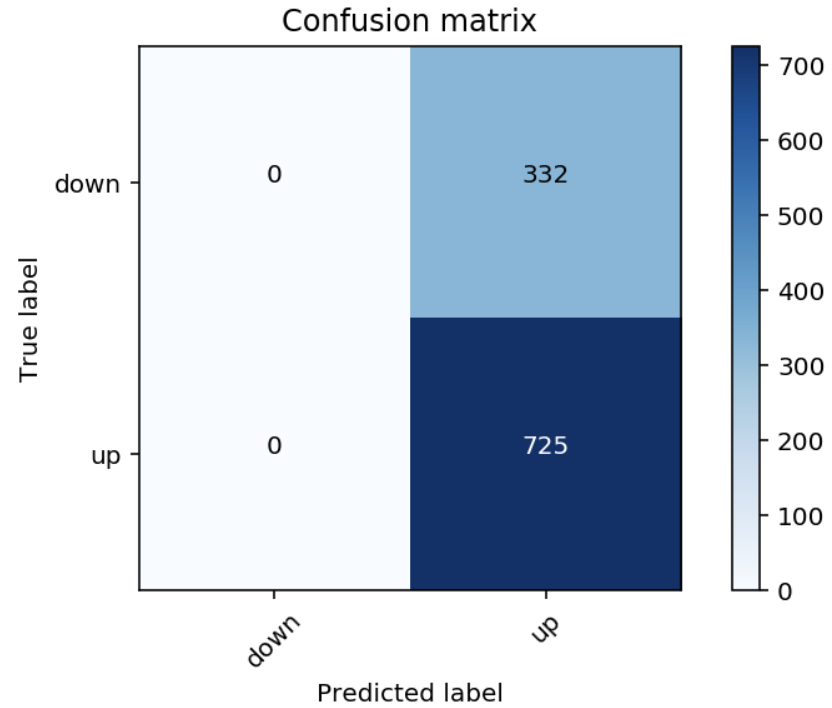}
	  \label{fig:confusion-matrix}
\end{figure}

Figure \ref{fig:confusion-matrix} illustrates this effect for the following model applied
to SPY daily data.

\begin{table}[!h]
	\centering	
	\begin{tabular}{|l|r|}
		\hline
	    Lookback (Slices) & 20 bars \\
		Features & Close  \\
		Label &  20 bars up/down \\	
		1 LSTM &  20 units (tanh) \\
		1 Dense output &  2 units (softmax) \\
		Bias & No \\
		Dropout & No \\
		Recurrent Dropout & No \\
		Optimizer & Adam \\
		Epochs & 100 \\
		Batch Size & 64 \\
		\hline
	\end{tabular}
	\caption{Model Detail and Hyperparameters}\label{table:ads-test}
\end{table}

\newpage

The model only predicts the \textit{Up} label, resulting in a validation precision of 0.6859 that correspond 
to the class imbalance as seen in the confusion matrix.

The simplest way to fight this consists on downsampling the \textit{Up} labels on the training set in order to get a perfectly even class balance. 

Retraining the same model results in a different confusion Matrix not only predicting the Upside anymore.

\begin{figure}[!h]
	\caption{Confusion Matrix on Validation Set}
	\centering
	  \includegraphics[width=0.45\textwidth]{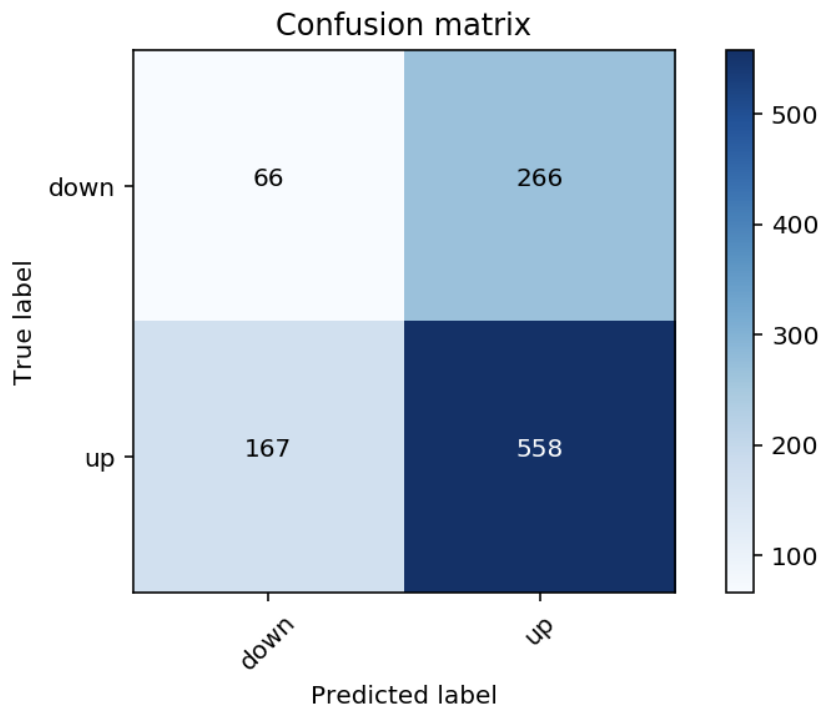}
	  \label{fig:confusion-matrix-balanced}
\end{figure}

Note that intraday data are less subject to such bias.

\newpage
\section{Conclusion}

This article covered the specificity of Financial Time Series data and the processing practice that can be 
applied to start a research project in good conditions.

It explains how to properly scale, slice and test the stationarity of the dataset.

It also speaks about features and proposed different labeling methods.

Time Series are not like any other data. The classical pre-processing methods generally used
cannot directly be applied. 

On top of that, their Financial nature means they follow a stochastic process, which is even adding another level of complexity.

The methods shown in this document can be applied to any financial instrument (Equities, Commodities, Forex, ...) and to 
any timescales (daily, hourly, 5 minutes, ...).

Selecting features, scaling and labelling is part the whole Machine Learning Research process.
Comparing different combinations of features and labels can sometimes have more 
impacts on the results than hyperparameters tuning.


\begin{thebibliography}{}

\bibitem{IntroTSF}
Peter J. Brockwell \& Richard A. Davis (2002) \\
\textit{Introduction to Time Series and Forecasting, Second Edition}


\bibitem{Forecasting} 
Rob J Hyndman \& George Athanasopoulos (2018) \\
\textit{Forecasting: Principles and Practice, Second Edition}
\\\texttt{https://otexts.com/fpp2/}

\bibitem{statFinTS} 
Kjersti Aas \& Xeni K. Dimakos (2004) \\
\textit{Statistical modelling of financial time series: An introduction}

\bibitem{analysisTS} 
Ruey S. Tsay (2010) \\
\textit{Analysis of Financial Time Series}

\bibitem{AFML} 
M. L. De Prado (2018) \\
\textit{Advances in Financial Machine Learning}

\bibitem{DF}
D. A. Dickey \& W. A. Fuller (1979) \\
\textit{Distribution of the Estimators for Autoregressive Time Series with a Unit Root}


\bibitem{introStatTS}
W. A. Fuller (1996) \\
\textit{Introduction to Statistical Time Series}


\bibitem{ADF}
Wikipedia. \\
\textit{https://en.wikipedia.org/wiki/Augmented\_Dickey–Fuller\_test}



\end{thebibliography}
\end{document}